\documentstyle[prl,aps,floats,epsf]{revtex}
\newcommand{\biblabel}[1]{[#1]} %
\renewcommand{\references}{%
\ifpreprintsty
\newpage
\else
\vskip3pt
\hrule width\hsize\relax
\vskip -0.2in
\fi
\list{\biblabel{\arabic{enumiv}}}%
{\labelwidth\WidestRefLabelThusFar  \labelsep4pt %
\leftmargin\labelwidth %
\advance\leftmargin\labelsep %
\ifdim\baselinestretch pt>1 pt %
\parsep  4pt\relax %
\else %
\parsep  0pt\relax %
\fi
\itemsep\parsep %
\usecounter{enumiv}%
\def\theenumiv{\arabic{enumiv}}%
}%
\let\newblock\relax %
\sloppy\clubpenalty4000\widowpenalty4000
\sfcode`\.=1000\relax
\ifpreprintsty\else\small\fi
}
\newcommand{\smeq}{\! = \!}

\newcommand{\smpl}{\! + \!}
\newcommand{\smmi}{\! - \!}
\newcommand{\Pintra}{P_{\rm intra}}
\newcommand{\Pinter}{P_{\rm inter}}
\begin{document}
\draft

\twocolumn[{   %% start twocolumn, but keep abstract single column

\title{
Interactions and Interference in Quantum Dots:\\
Kinks in Coulomb Blockade Peak Positions
}

\author{Harold U. Baranger,$^{1}$ Denis Ullmo,$^{2}$ and 
Leonid I. Glazman$^{3}$}

\address{$^{1}$Bell Laboratories-- Lucent Technologies,
700 Mountain Ave., Murray Hill NJ 07974}
\address{$^{2}$Laboratoire 
de Physique Th\'eorique et Mod\`eles Statistiques (LPTMS), 
91405 Orsay Cedex, France}
\address{$^{3}$Theoretical Physics Institute, University of Minnesota,
Minneapolis MN 55455}

\date{ 
5 May 1999; revised 16 July 1999
}

\maketitle

\mediumtext
\vspace*{-0.2 truein}
\begin{abstract}
We investigate the spin of the ground state of a geometrically confined
many-electron system. For atoms, shell structure simplifies this problem--
the spin is prescribed by the well-known Hund's rule. In
contrast, quantum dots provide a controllable setting for studying the
interplay of quantum interference and electron-electron interactions in 
general cases. In a generic confining potential, the shell-structure
argument suggests a singlet ground state for an even number of electrons.
The interaction among the electrons produces, however, accidental
occurrences of spin-triplet ground states, even for weak interaction, a
limit which we analyze explicitly. Variaton of an external parameter causes
sudden switching between these states and hence a kink in
the conductance. Experimental study of these kinks would yield the exchange
energy for the ``chaotic electron gas''.
\end{abstract}
\vspace*{0.15 truein}
%\pacs{PACS numbers: 05.45.+b, 73.20.Dx, 73.23.-b, 73.23.Hk, 73.40.Gk}
%\vspace*{-0.1 truein}

}]  %% end single column part, start twocolumn formatting

\narrowtext

The evolution of the properties of a system as a continuous change is made
to it is a ubiquitous topic in quantum physics. The classic example is the
evolution of energy levels as the strength of a perturbation is varied
\cite{Merzbacher}. Typically, neighboring energy levels do not cross each
other, but rather come close and then repel in an ``avoided crossing''.
However, if there is an exact symmetry, neighboring levels can have
different symmetries uncoupled by the perturbation, and in this special case
they can cross.

A new and powerful way of studying parametric evolution in many-body quantum
systems is through the Coulomb blockade peaks that occur in mesoscopic
quantum dots \cite{McEuen,Ashoori,MarcusStewart97,Kouw97}. The electrostatic
energy of an additional electron on a quantum dot-- an island of confined
charge with quantized states-- blocks the flow of current through the dot--
the Coulomb blockade \cite{CblockNato,MesoNato}. Current can flow only if
two different charge states of the dot are tuned to have the same energy;
this produces a peak in the conductance through the dot. The position of the
Coulomb blockade peak depends on the difference in ground state energy upon
adding an electron, $E_{\rm gr}(N \smpl 1) \smmi E_{\rm gr}(N)$. Thus, the
evolution of the peak position reflects the evolution of these many-body
energy levels as a parameter, such as magnetic field or gate voltage, is
varied.

Since quantum dots are generally irregular in shape, the orbital levels have
no symmetry and so avoid crossing. However, the spin degrees of freedom are
often decoupled from the rest of the system, so states of different total
spin can cross. Such a crossing will cause an abrupt change in the evolution
of the Coulomb blockade peak position-- a kink-- as the spin of the ground
state changes, even though there is no crossing in the single-particle
states. Such kinks have been observed in small circular dots-- ``artificial
atoms''-- because of their special circular symmetry \cite{Kouw97}.
More generally, kinks should occur in generic quantum dots
with no special symmetries \cite{AndKam98}. In fact, the data of Ref.
\cite{MarcusStewart97} show evidence for kinks in large dots, though they
were not the subject of that investigation.

Small circular dots behave much like atoms (hence ``artificial atoms''): the
circular symmetry causes degeneracy of the orbital levels and so
a large spacing between allowed energies. In sharp contrast, there is no
degeneracy in irregular dots: the typical single-particle orbital level separation
is simply given by $\Delta \equiv 1/\nu V$ where $\nu$ is the bulk density
of states and $V$ is the volume of the dot. Kinks in the evolution of
Coulomb blockade peak positions may occur whenever the ground state of the
dot is separated from an excited state with different spin by an energy of
order $\Delta$. The interference effects causing the separation are unique to
each state and change upon tuning. In fact, the two states may switch at a
certain point, the former excited state becoming the ground state: such
switching corresponds to a kink. Note that kinks occur in pairs: kinks in the
peaks corresponding to the $N \!  \rightarrow \! N \smpl 1$ transition and
that for $N \smmi 1 \! \rightarrow \! N$ both occur when $E_{\rm gr} (N)$
switches. We see from this argument that kinks in the evolution of peak
positions with parameter is a general property of quantum dots \cite{AndKam98}.

Here we analyze these kinks explicitly in a particular
limit.  Consider a large quantum dot in which the single-particle properties
are ``random'': the statistics of the energies follow the classic random
matrix ensembles \cite{Mehta} and the wave-functions obey Gaussian statistics
with a correlation function given by the superposition of random plane waves
\cite{BerrySrednicki}. The single-particle properties of such random systems
have been extensively investigated, and it has been conjectured, with
considerable evidence, that these are good models for quantum dots in which
the classical dynamics is chaotic \cite{MesoNato,BarWestRev}.
%Thus in this limit a quantum dot contains a ``chaotic electron gas''.

To treat the Coulomb blockade, we must consider not only single-particle
properties but interactions among the particles as well. One natural way to
proceed is to treat these interactions in the basis of self-consistent
single-particle states $\{ \psi_m ({\bf r})\chi_\sigma(s) \}$, where $m$ and
$\sigma$ are the labels of orbital and spin states respectively (we neglect
the weak spin-orbit interaction). In the limit of zero interaction, two
electrons occupy each filled orbital state, except for the top level when the
total number of electrons is odd.  Because of the interference produced by
confinement in the dot, the electron density is not smooth but rather has
small deviations from the classical-liquid result. Due to the
electron-electron interaction, these deviations contribute to the
ground-state energy, in addition to the conventional ``classical-liquid''
charging energy. If the interaction does not change the double-occupancy of
the levels, one finds that the part of such contribution coming from the last
doubly-occupied level $n$ is
\begin{eqnarray}
   \xi_{n,n} & \equiv & \int \! d{\bf r} d{{\bf r}'}
   [ |\psi_n ({\bf r})|^2 - \langle |\psi_n ({\bf r})|^2 \rangle ]
   V_{\rm scr} ({\bf r}-{\bf r}') 
   \nonumber \\
   && \times [ |\psi_n ({\bf r}')|^2 - \langle |\psi_n ({\bf r}')|^2 \rangle ] \, .
   \label{xinm}
\end{eqnarray} 
It is appropriate to use the short-ranged screened interaction $V_{\rm scr}$
here since the smooth background of the other electrons provides screening;
$\langle\dots\rangle$ denotes the standard ensemble averaging.  If, because
of the interactions, one of the electrons of that level is promoted to the
next orbital state, the result (\ref{xinm}) is modified to become
\begin{eqnarray}
   &&\xi_{n,n+1}^\pm \equiv \int \! d{\bf r} d{{\bf r}'}
   [ |\psi_n ({\bf r})|^2 - \langle |\psi_n ({\bf r})|^2 \rangle ]
   V_{\rm scr} ({\bf r}-{\bf r}') 
   \nonumber \\
   && \qquad \qquad \times 
[ |\psi_{n+1} ({\bf r}')|^2 - \langle |\psi_{n+1} ({\bf r}')|^2 \rangle ] 
   \label{pm}
\\
&&\qquad \pm \int \! d{\bf r} d{\bf r}' 
  \psi_n ({\bf r})\psi_{n+1}^* ({\bf r}) V_{\rm scr} ({\bf r}-{\bf r}') 
\psi_n^*({\bf r}')\psi_{n+1} ({\bf r}') \, . \nonumber
\end{eqnarray} 
The signs $+$ and $-$ in Eq.~(\ref{pm}) correspond to the singlet and
triplet states respectively.

For a large ballistic dot, $k_FL\gg 1$, the lack of correlation among the
random wavefunctions $\psi_n$ and $\psi_m$ with $n\neq m$ leads to a
hierarchy of the matrix elements of the interaction\cite{matrixels} (here
$k_F$ is the Fermi wave vector of electrons in the dot, and $L$ is the linear
size of the dot). The first integral in Eq.~(\ref{pm}) vanishes in the limit
$k_FL\to\infty$, and one is left only with the second, exchange interaction,
term. In the same limit, the exchange interaction term has a non-zero average
value and vanishingly small mesoscopic fluctuations. The lowest of two
energies $\xi_{n,n+1}^\pm$ corresponds to the triplet state ($\pm\to -$).
Neglecting the small mesoscopic fluctuations of the energies $\xi_{n,n}$ and
$\xi_{n,n+1}^-$, one finds the difference between the energies of the singlet
state formed by doubly-occupied levels and of the triplet state with two
singly-occupied orbital levels:
\begin{eqnarray}
\xi&\equiv&\langle\xi_{n,n}\rangle -\langle\xi_{n,n+1}^-\rangle
\label{defxi}
\\
&=&2\int \! d{\bf r} d{\bf r}' V_{\rm scr} ({\bf r}-{\bf r}')
\left|F({\bf r}-{\bf r}')\right|^2
= 2 \overline{ \hat{V}_{\rm scr} ({\bf k}-{\bf k}') } \,.
\nonumber 
\end{eqnarray} 
Here $F({\bf r})\equiv\overline{\exp (i{\bf k}'{\bf r})}$, with the bar
denoting the average over the Fermi surface $|{\bf k}'|=k_F$.

In the above argument we have implicitly assumed the absence of
time-reversal symmetry. For $B \smeq 0$, the same basic argument holds,
but the interaction in the Cooper channel should be included; this 
increases $\xi$ making the proportionality constant in Eq. (3) larger
than 2.

We thus consider a model \cite{BlaMirMuz} with a single non-zero interaction
constant $\xi$. This quantity is simply related to the usual electron gas
parameter $r_s$ (the ratio of the Coulomb energy at the mean interparticle
distance to the Fermi energy): $\xi \approx (1/\sqrt{2}\pi)
r_s\ln(1/r_s)\Delta$ for small $r_s$ and in the absence
of time-reversal symmetry. As $r_s$ increases, the considerations
discussed here apply until $\xi$ exceeds $\Delta$ at which point the Stoner
criterion for a magnetic instability is approached. For instance, for $r_s
\smeq 1$ averaging the Thomas-Fermi screened potential over the Fermi surface
yields $\xi \smeq 0.6 \Delta$ in two dimensions. 

The distribution of electrons among the levels depends on the single particle
level spacing compared to $\xi$. This is particularly clear when the total
number of electrons $N$ is even: the top two electrons can either be in the
same orbital level at a cost of $\xi$ in interaction energy or one can be in
level $N/2$ and the other in $N/2+1$ at an energy cost of $\epsilon_{N/2+1} -
\epsilon_{N/2}$. Since the magnitudes of both $\xi$ and $\epsilon_{N/2+1} -
\epsilon_{N/2}$ are of order $\Delta$, sometimes the top level is doubly
occupied and sometimes not. In the case of double occupation, the state is,
of course, a singlet; if two sequential levels are occupied, the exchange
interaction leads to a triplet state. If at most two orbital levels are
singly occupied, the ground state energy of a dot is, then, a sum of three terms:
\begin{equation}
   E_{\rm gr} = E_{\rm ch} +
   \sum_{(n,\sigma)~{\rm filled}} \epsilon_{n,\sigma} +
   M \xi
   \label{Eground}
\end{equation}
where $M$ is the number of doubly occupied levels. For our arguments, the
number of electrons is constant and so the charging energy $E_{\rm ch}$ is
irrelevant. Note that the energy (\ref{Eground}) is equivalent to the
simultaneous filling of two sequences of levels, one of which is rigidly
shifted by $\xi$ from the other \cite{BlaMirMuz}. Finally, if several
orbital level spacings in sequence are small, more complicated configurations
occur for both even and odd $N$ \cite{BrouOregHalp,Aleiner}. Moreover, the
problem of the ground-state spin of a mesoscopic system becomes very
complicated upon approaching the Stoner instability \cite{Aleiner}.

\begin{figure}[t]
\begin{center}
\leavevmode
\epsfxsize = 7.0cm
\epsfbox{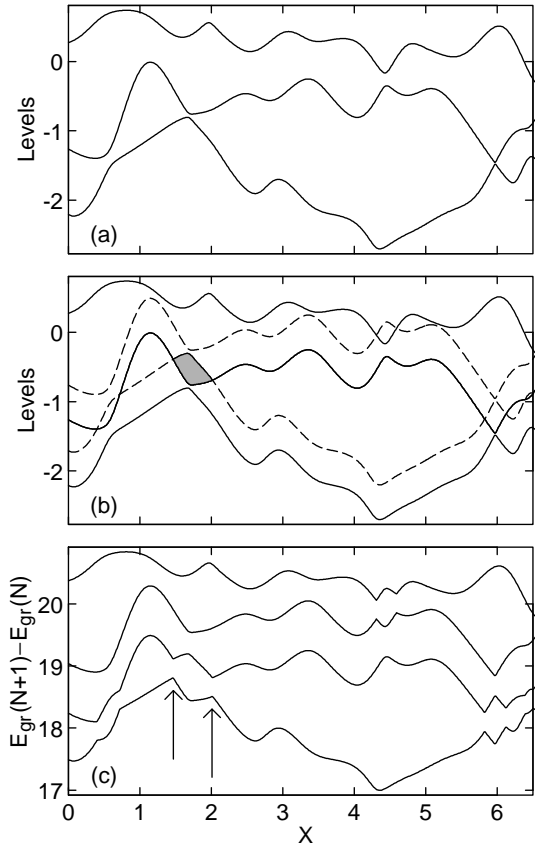}
\end{center}
\caption{
Mechanism for kinks in the Coulomb blockade peak positions:
Gaussian process simulation results as a function of scaled
parameter $X$ ($\beta \smeq 2$).
{\bf (a)} 3 typical orbital energy levels; note the infrequent avoided
crossings. 
{\bf (b)} Two sets of energy levels, one for singly occupied
orbitals (solid) and the second for double occupancy (dashed). For a large
dot, the doubly occupied levels are rigidly shifted from those for single
occupancy by the interaction $\xi$ ($=0.5$ here), see
Eq.~(\protect\ref{Eground}). Consider an even number of electrons in which
the top two particles are placed in the levels shown. When the
doubly-occupied state (first dashed line) has higher energy than the next
orbital level (second solid line), the ground state has two partially
occupied levels-- a triplet spin state. When these two levels cross, the
lowest energy in the triplet sector crosses the lowest singlet state. 
{\bf (c)} Difference in energy upon adding an electron, proportional to the
Coulomb blockade peak position. (Traces offset by 0.3 for clarity.)
Crossings of the singlet and triplet levels cause kinks in the peak positions
(arrows mark 2 examples). Two adjacent peaks are affected by each change in a
ground state: they have kinks at the same $X$ and switch which orbital level
they are tracking.
} 
\end{figure}

As a parameter is varied, the single particle energies change and may cause
a change in the level occupations and so a kink. This is explained
and illustrated in Fig.~1. 

The distribution of the kinks in the parameter space follows from a random
matrix model. We assume that the Hamiltonian of the dot follows the Gaussian
Orthogonal Ensemble in the absence of a magnetic field (GOE, $\beta \smeq 1$) 
and the Gaussian Unitary Ensemble for nonzero $B$ (GUE, $\beta \smeq 2$)
\cite{Mehta}. The dependence on the parameter $X$ is included by means of a 
Gaussian process \cite{Wilkinson,Alhassid95};
we consider the process
\begin{equation}
   H(x) = H_1 \cos(X) + H_2 \sin(X)
   \label{Gprocess}
\end{equation}
where the distributions of $H_1$ and $H_2$ follow the appropriate Gaussian
ensemble. Extensive work on parametric correlations has shown that the
properties of Gaussian processes are universal when the energy is measured in
units of the mean level separation $\Delta$ and the parameter is scaled by
the typical rate at which energies are perturbed, $\langle (d\epsilon /dX)^2
\rangle^{1/2}$ \cite{Wilkinson,Alhassid95,AltSimonsRev}.  For simplicity we
restrict our attention to kinks caused by a change in configuration of the
top two electrons when $N$ is even.

The first quantity to consider is the mean density of kinks, $\rho_{\rm
kink}$. First, because a kink occurs when $\epsilon_{N/2+1} - \epsilon_{N/2}
\smeq \xi$, this density is proportional to the probability of having such a
level-separation. Second, $\rho_{\rm kink}$ must reflect the rate at which
the levels change as a function of parameter. In fact, it is known that the
distribution of the slopes of the levels, $d\epsilon /dX$, is Gaussian and
independent of the distribution of the levels themselves \cite{Wilkinson}.
Thus the two contributions are simply multiplied:
\begin{eqnarray}
   \rho_{\rm kink} & = & \frac{2}{\sqrt{\pi}} \,
   p_{\epsilon}^{(\beta)} (\xi) 
   \left\langle ( \frac{d\epsilon}{dX} )^2 \right\rangle^{1/2}
%   {\rm rms}\! \left( \frac{d\epsilon}{dX} \right) \,
   \label{meandens}
   \\[0.10in]
   & \propto & \xi^{\beta} \quad \mbox{for small $\xi$}
   \nonumber
\end{eqnarray}
where $p_{\epsilon}^{(\beta)} (s)$ is the distribution of nearest-neighbor
level separations, for which the Wigner surmise \cite{Mehta} is an excellent
approximation. $\rho_{\rm kink}$ has a strong dependence on $\beta$ when
$\xi$ is small because of the symmetry dependence of
$p_{\epsilon}^{(\beta)}$, and so will be sensitive to a magnetic field. In
fact, {\it the sensitivity to magnetic field could be used to extract
experimental values for $\xi$ in quantum dots-- a direct measure of the
strength of interactions.}

Next, an important property is the spacing in $X$ between two neighboring
kinks.  For $\xi \! \ll \! \Delta$, kinks occur when two orbital levels come
very close and so are caused by avoided crossings of single-particle levels.
Since each avoided crossing produces two kinks, kinks will occur in pairs,
with small intra-pair and large inter-pair separations.  The behavior near an
avoided crossing is dominated by just two levels and characterized by 3
parameters-- the mean and difference of the slopes far from the crossing and
the smallest separation.  Wilkinson and Austin have derived the joint
probability distribution of these parameters for Gaussian random processes
\cite{Wilkinson}. By expressing the intra-pair separation in terms of these
parameters and integrating over the joint distribution, we find that the
distribution of intra-pair separations is
\begin{eqnarray}
   \lefteqn{
   \Pintra (x)  = 
   2 \frac{x}{\xi^2} \int_{\textstyle 0}^{\textstyle \xi^2/x^2}
   \!\!\! du \, \exp (-u) 
   } \\
   & & \times \left\{  
   \begin{array}{ll}
      u^2\,, & \, \beta=2 \\
      u^{3/2} (1-ux^2/\xi^2 )^{-1/2} /\sqrt{\pi}\,, & \, \beta=1
   \end{array}
   \right.  . \nonumber
\end{eqnarray}
For small $x$, $\Pintra$ is linear in the separation $x$ both with and
without a magnetic field.

The separation between pairs is usually large for small $\xi$--
avoided crossings with small gap are rare--
typically many correlation lengths.
Hence there is no correlation between pairs: $\Pinter (x)$ 
is Poisson (exponential) for large $x$. 
Correlation suppresses the probability of two close crossings.
We make a simple model for this by assuming $\Pinter \propto x$ for
$x < x_0$ and so approximate $\Pinter$ by
\begin{equation}
   \Pinter (x) = \frac{C}{\alpha}
   \left\{
   \begin{array}{ll}
       \exp(-(x-x_0)/\alpha )\,, & \, x > x_0 \\
        x/x_0 \,, & \, x < x_0
   \end{array}
   \right.
\end{equation}
where $C$ is for normalization.
$x_0$ should be of order 1 in scaled units; we choose it to be 
the minimum of the correlation function of $d\epsilon /dX$,
$x_0 \smeq 0.85$ ($0.6$) for GOE (GUE).
The mean density Eq. (\ref{meandens}) sets alpha,
\begin{equation}
   1 / \rho_{\rm kink} = \langle x \rangle = 
   \case{1}{2} [ \langle x \rangle_{\rm intra} + 
   \langle x \rangle_{\rm inter} ] 
   \, ,
\end{equation}
combined with the distribution $\Pintra$. 
In this way we have a parameter free expression for the distribution 
of the separation between adjacent kinks.

\begin{figure}[tb]
\begin{center}
\leavevmode
\epsfxsize = 7.0cm
\epsfbox{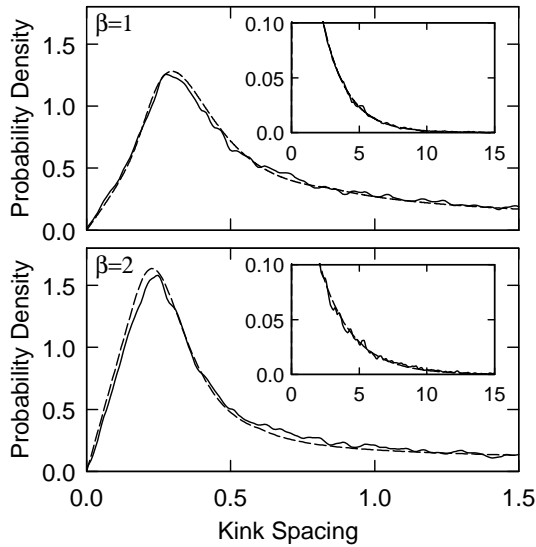}
\end{center}
\caption{
Distribution of the separation between nearest-neighbor kinks in the
Coulomb blockade peak positions. The interaction strength is $\xi=0.5$. 
{\bf Top:} zero magnetic field (time-reversal symmetry, $\beta=1$). 
{\bf Bottom:} non-zero magnetic field ($\beta=2$). 
{\bf Insets:} tails of the distributions.
Both the analytic theory for small $\xi$ (dashed line) and the results of
Gaussian process simulations (solid line) are shown. The excellent agreement
is remarkable considering that $\xi=0.5$ is not small and the absence of any
fitting parameter.
} 
\end{figure}

While the above theory is for $\xi \! \ll \! \Delta$, in the experiments 
\cite{MarcusStewart97,Kouw97} $r_s \! \sim \! 1$ so that 
$\xi \! \lesssim \! \Delta$. Hence we turn to numerical calculation to test
the range of validity of our expressions. Gaussian processes were produced
using the Hamiltonian (\ref{Gprocess}) with matrix size $200$ over the full
interval $X \in [0,2\pi]$; the middle third of the spectrum was used. Sample
energy levels for $\beta \smeq 2$ are shown in Fig. 1.

Fig. 2 shows the distribution of kink spacings for the experimentally
relevant value $\xi \smeq 0.5$ and $\beta \smeq 1, \, 2$. Though
outside the regime of immediate applicability, the theoretical
curves agree closely with the numerical data. Thus the simple
two-level calculation captures the main features of the kink distribution
for $\xi \! \lesssim \! 1$ and so is adequate for describing experiments
in large dots \cite{MarcusStewart97}.

In conclusion, through the properties of kinks in the Coulomb blockade
traces, experiments on quantum dots can directly determine the main
interaction parameter in these dots and so obtain the exchange energy for the
chaotic electron gas.

We thank C.M. Marcus for a stimulating conversation which helped initiate
this work and acknowledge valuable discussions with K.A. Matveev and I.L.
Aleiner. HUB and LIG appreciate the hospitality of the ICCMP, Brasilia
Brazil, where this work started. 
After completion of this work, we learned of Ref. \onlinecite{BrouOregHalp}
by P.W. Brouwer, Y. Oreg, and B.I. Halperin
in which some similar results were obtained.
Finally, we acknowledge support of NSF Grant DMR-9731756; 
the LPTMS is ``Unit\'e de recherche de l'Universit\'e Paris~11 
associ\'ee au C.N.R.S.''.


\begin{references}


\bibitem{Merzbacher}
E. Merzbacher, {\it Quantum Mechanics} (Wiley, New York, 1970) pp. 428-9;
A.A. Shapere and F. Wilczek, Eds., {\it Geometric Phases in Physics}
(World Scientific, Singapore, 1989).

\bibitem{McEuen}
P.L. McEuen, E.B. Foxman, U. Meirav, M.A. Kastner, Y. Meir, 
N.S. Wingreen, and S.J. Wind, Phys. Rev. Lett. {\bf 66}, 1926 (1991).

\bibitem{Ashoori}
R.C. Ashoori, H.L. Stormer, J.S. Weiner, L.N. Pfeiffer, S.J. Pearton,
K.W. Baldwin, and K.W. West, Phys. Rev. Lett. {\bf 68}, 3088 (1992).

\bibitem{MarcusStewart97}
D.R.~Stewart, D.~Sprinzak, C.M.~Marcus, K.~Campman and A.C.~Gossard, 
Science {\bf 278}, 1784 (1997).

\bibitem{Kouw97}
L.P. Kouwenhoven, T.H. Oosterkamp, M.W.S. Danoesastro, M. Eto, 
D.G. Austing, T. Honda, and S. Tarucha, 
Science {\bf 278}, 1788 (1997).

\bibitem{CblockNato}
H.~Grabert and M.H.~Devoret,
{\it Single Charge Tunneling: Coulomb Blockade Phenomena in Nanostructures}
(Plenum Press, New York, 1992).

\bibitem{MesoNato} 
L.P. Kouwenhoven, {\it et al.} in {\it Mesoscopic Electron Transport},
L.L.~Sohn, L.P.~Kouwenhoven, and G.~Sch\"on, Eds.
(Kluwer, Dordrecht, 1997) pp. 105-214.

\bibitem{AndKam98}
A.V. Andreev and A. Kamenev,
Phys. Rev. Lett. {\bf 81}, 3199 (1998).

\bibitem{Mehta}
M.L. Mehta, {\it Random Matrices} (Academic, New York, 1991).

\bibitem{BerrySrednicki}
M.V. Berry, J. Phys. A {\bf 10}, 2083 (1977);
M. Srednicki, Phys. Rev. E {\bf 54}, 954 (1996).

\bibitem{BarWestRev}
H.U. Baranger and R.M. Westervelt, in {\it Nanotechnology}, 
G. Timp, Ed. (Springer-Verlag, New York, 1999) pp. 537-628.

\bibitem{matrixels}
Ya.M. Blanter, Phys. Rev. B {\bf 54}, 12807 (1996);
O. Agam, N.S. Wingreen, B.L. Altshuler, D.C. Ralph, and M. Tinkham, 
Phys. Rev. Lett, {\bf 78}, 1956 (1997); 
B.L. Altshuler, Y. Gefen, A. Kamanev, and L.S. Levitov,
Phys. Rev. Lett. {\bf 78}, 2803 (1997);
I.L. Aleiner and L.I. Glazman, Phys. Rev. B {\bf 57}, 9608 (1998).

\bibitem{BlaMirMuz}
Ya.M. Blanter, A.D. Mirlin, and B.A. Muzykantskii, 
Phys. Rev. Lett. {\bf 78}, 2449 (1997).

\bibitem{BrouOregHalp}
P.W. Brouwer, Y. Oreg, and B.I. Halperin, 
cond-mat/9807148.

\bibitem{Aleiner}
I. Kurland, I.L. Aleiner, and B.L. Althsuler, unpublished.

\bibitem{Wilkinson}
M. Wilkinson, J. Phys. A {\bf 22}, 2795 (1989);
E.J. Austin and M. Wilkinson, Nonlinearity {\bf 5}, 1137 (1992).

\bibitem{Alhassid95}
H. Attias and Y. Alhassid, Phys. Rev. B {\bf 52}, 4776 (1995).

\bibitem{AltSimonsRev}
B.L. Altshuler and B.D. Simons, in {\it Mesoscopic Quantum Physics},
E. Akkermans, G. Montambaux, J.-L. Pichard, and J. Zinn-Justin, Eds.
(North-Holland, New York, 1995) pp. 1-98.

\end{references}
\end{document}